\documentclass[10pt,letterpaper]{article}
\usepackage{li}
\usepackage{graphicx,psfrag,bm,amsmath,amsfonts,amssymb,color,subfigure}

\newcommand{\be}{\begin{equation}}
\newcommand{\ee}{\end{equation}}
\newcommand{\bea}{\begin{eqnarray}}
\newcommand{\eea}{\end{eqnarray}}

\newcommand{\eps}{\epsilon}

\newcommand{\re}[1]{\mbox{Re}\left[#1\right]}
\newcommand{\g}{\gamma_{\perp}}
\newcommand{\gpar}{\gamma_{\|}}
\newcommand{\bx}{\bm{x}}
\newcommand{\bxp}{\bm{x}^{\prime}}
\newcommand{\tk}{\tilde{k}}

\begin{document}
\title{Quantitative Verification of Ab Initio Self-consistent Laser Theory}

\author{ Li Ge,$^1$ Robert J. Tandy,$^1$  A. Douglas
Stone,$^{1}$  and Hakan E. T\"ureci$^2$ }

\address{ $^1$Department of Applied Physics, P. O. Box 208284, Yale
University, New Haven, CT 06520-8284, USA \\
$^2$Institute of Quantum Electronics, ETH Zurich, 8093 Zurich,
Switzerland}

\begin{abstract}
We generalize and test the recent ``ab initio'' self-consistent
(AISC) time-independent semiclassical laser theory.  This
self-consistent formalism generates all the stationary lasing
properties in the multimode regime (frequencies, thresholds,
internal and external fields, output power and emission pattern)
from simple inputs: the dielectric function of the passive cavity,
the atomic transition frequency, and the transverse relaxation time
of the lasing transition. We find that the theory gives excellent
quantitative agreement with full time-dependent simulations of the
Maxwell-Bloch equations after it has been generalized to drop the
slowly-varying envelope approximation.  The theory is infinite order
in the non-linear hole-burning interaction; the widely used third
order approximation is shown to fail badly.
\end{abstract}



\noindent The Maxwell-Bloch (MB) equations provide the foundation of
semiclassical laser theory \cite{haken1985} and are the simplest
description which captures the full space-dependent non-linear
behavior of a laser. These time-dependent equations can be simulated
to determine the stationary lasing state.  However {\it
time-independent} methods to find these stationary properties in the
multi-mode regime for an open laser cavity did not exist until the
recent work of Tureci {\it et al.}
\cite{tureci2006,tureci2007,tureci2008} presented an ``ab initio''
self-consistent (AISC) formalism which generates all of the lasing
properties including the output power and emission pattern from a
few simple inputs. The laser cavity can be of arbitrary complexity
and openness, including, e.g., chaotic dielectric disk lasers
\cite{gmachl1998}, photonic lattice defect mode lasers
\cite{painter1999} and random lasers \cite{tureci2008,cao2005}. Here
we generalize this infinite order non-linear theory by extending it
beyond the slowly-varying envelope approximation.  With this
improvement it gives remarkably good agreement with time-dependent
simulations of the Maxwell-Bloch (MB) equations, while the standard
third order approximation to the non-linear hole-burning interaction
fails badly.

The AISC theory builds on the original ideas of Haken and coworkers
\cite{haken1963,fu1991} that the inversion of the lasing medium will
be approximately time-independent when $\gpar \ll \g,\Delta$ (where
$\g$ is the transverse (polarization) relaxation rate, $\Delta$ is
the typical mode spacing, and $\gpar$ is the longitudinal
(inversion) relaxation rate).  The only significant approximation in
the theory presented below is this approximation of stationary
inversion (SIA) (well-satisfied in many lasers of interest). In
addition to the excellent agreement we find between the AISC theory
and the MB simulations when the ratios $\gpar/\g,\gpar/\Delta$ are
very small, we develop below a perturbative treatment of the beating
terms which are neglected in SIA to extend the theory to larger
values of these ratios. The key improvements contained in the AISC
theory are: 1) Treatment of the openness of the cavity exactly. 2)
Inclusion of the space-dependent non-linear modal interactions
(spatial hole-burning) to all orders, in contrast to standard
third-order treatments \cite{haken1985}.  We show below that the
third order treatment fails quantitatively and qualitatively even
for the simple laser resonator we study here.

\begin{figure}[htbp]
\centering
\includegraphics[width=7cm]{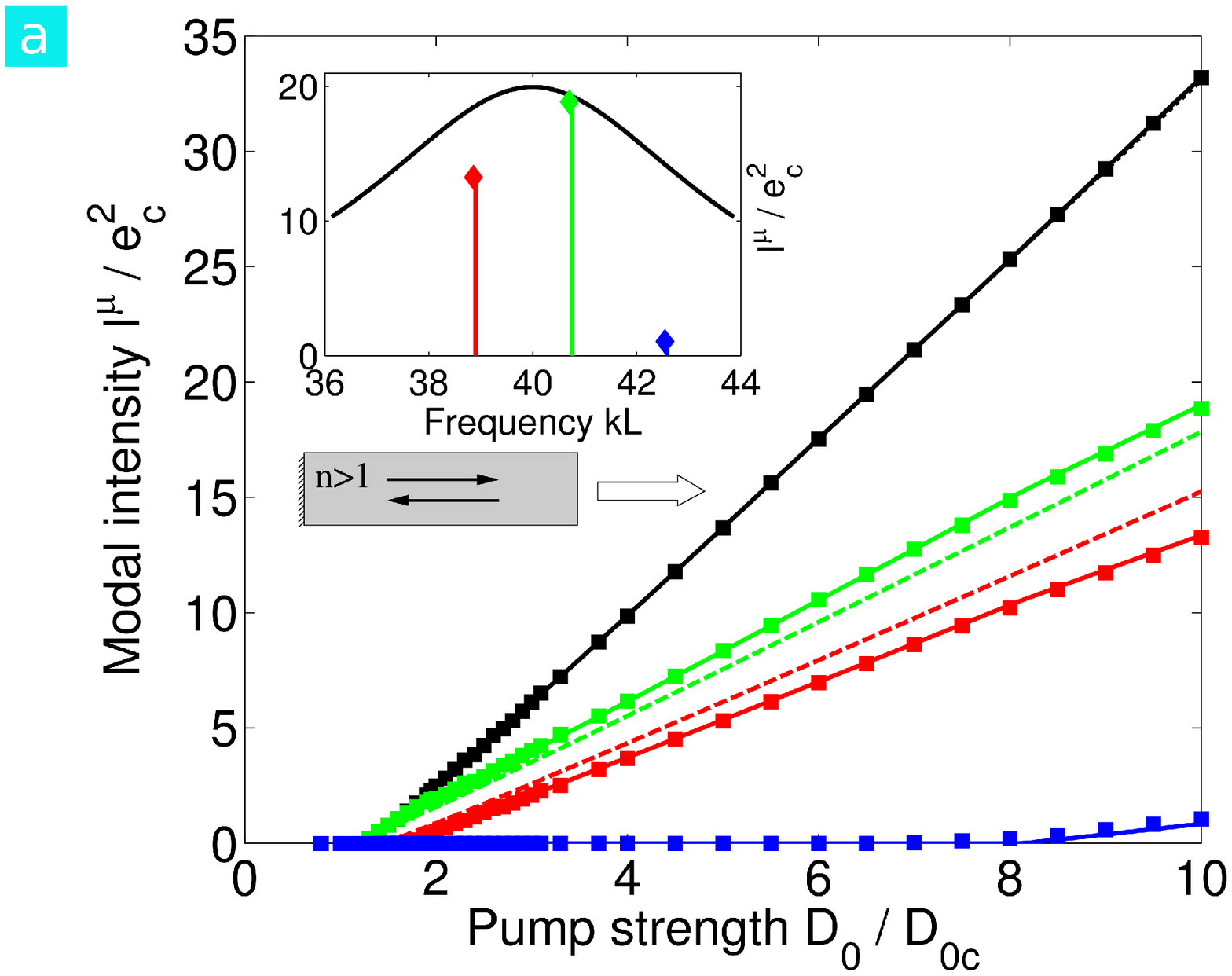}
\includegraphics[width=7cm]{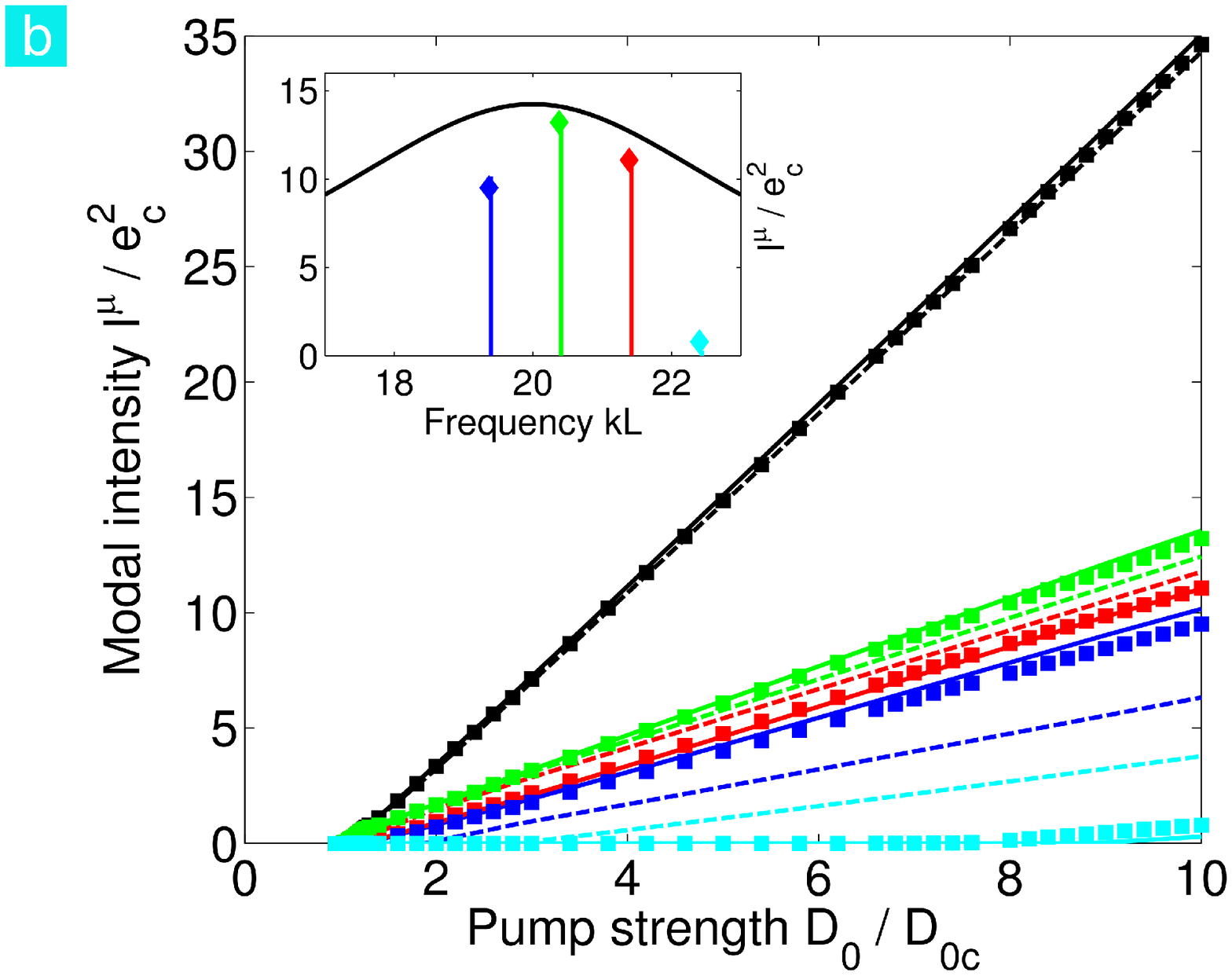}
\caption{Modal intensities as functions of the pump strength $D_0$
in a one-dimensional microcavity edge emitting laser of $\g=4.0$ and
$\gpar=0.001$. (a) $n=1.5$, $k_aL=40$. (b) $n=3$, $k_aL=20$. Square
data points are the result of MB time-dependent simulations; solid
lines are the result of time-independent ab initio calculations (AI)
of Eq. (\ref{eqTSG}).  Excellent agreement is found with no fitting
parameter.  Colored lines represent individual modal output
intensities; the black lines the total output intensity. Dashed
lines are results of AISC calculations when the slowly-varying
envelope approximation is
made as in Ref. [3] 
showing significant quantitative discrepancies. For example, in the
$n=3$ case the differences of the third/fourth thresholds between
the MB and AISC approaches are 46\%~/~63\%, respectively, but are
reduced to 3\% and 15\% once the SVEA is removed. The spectra at
$D_0=10$ and the gain curve are shown as insets in (a) and (b) with
the solid lines representing the predictions of the AISC approach
(Eq. (\ref{eqTSG})) and with the diamonds illustrating the height
and frequency of each lasing peak. The schematic in (a) shows a
uniform dielectric cavity with a perfect mirror on the left and a
dielectric-air interface on the right.} \label{figModalIntensity}
\end{figure}

To perform a well-controlled comparison of MB and AISC results we
chose to study the simple one-dimensional microcavity edge emitting
laser \cite{tureci2006,tureci2007} consisting of a perfect mirror at
the origin and a dielectric region of uniform index $n_0$ and length
$L$ terminating abruptly on air (see inset, Fig.
\ref{figModalIntensity}).  The MB equations are simulated in time
and space using a FDTD approach for the Maxwell equations, while the
Bloch equations are discretized using a Crank-Nicholson scheme.  To
avoid solving a nonlinear system of equations at each spatial
location and time step, we adopt the method proposed by
Bid\'{e}garay \cite{Bidegaray2003}, in which the polarization and
inversion are spatially aligned with the electric field, but are
computed at staggered times, along with the magnetic field.  Modal
intensities are computed by a Fourier transform of the electric
field at the boundary after the simulation has reached the steady
state.

The AISC theory consists of a set of coupled non-linear
time-independent integral equations of size equal to the number of
lasing modes at a
given pump. The AISC theory presented in Refs. [2, 3] 
is a solution to the MB equations \cite{haken1985} {\it after} two
standard simplifications, the rotating-wave approximation (RWA) and
slowly-varying envelope approximation (SVEA). The SVEA involves
factoring out the rapid time-dependence of the electric field and
the atomic polarization field at the atomic frequency, $k_a$, (here
and below we set $c=1$ and use frequency and wavevector
interchangeably), and neglecting the second time derivatives in the
Maxwell wave equation of the remaining envelope function of the
fields. The resulting non-linear system contains only first
time-derivatives of the fields and the inversion and is sometimes
referred to as the Schr\"odinger-Bloch (SB) equations
\cite{harayama1999}. The SVEA works well when the cavity frequencies
are negligibly shifted from the atomic frequency. For microcavities
these shifts need not be negligible; our simulated cavities have
$n_0 k_a L = 60$ corresponding to roughly ten wavelengths of
radiation inside the cavity, approaching the microcavity limit.  For
the case studied here, we find noticeable discrepancies between MB
solutions and the AISC theory of Refs. [2-4]
which incorporates the SVEA (see Fig. \ref{figModalIntensity}). This
motivated us to generalize the AISC theory to drop the SVEA (the RWA
is found to be well-satisfied in all cases).

The generalization was as follows. Again stationary periodic
solutions are assumed for the electric field $E(x,t) = 2 \re{
e(\bx,t)} = 2 \re{\sum_\mu \Psi_\mu(\bx) \exp{(-i k_\mu t)}}$ and
for the polarization fields, which oscillate at unknown lasing
frequencies, $k_\mu$.  The spatial variation of the field amplitude
$\Psi_\mu(\bx) $ is also unknown, and not assumed to be a cavity
resonance, but is determined self-consistently.  For a high finesse
cavity and near the first threshold it was shown \cite{tureci2006}
that $\Psi_\mu (\bx)$ is well-approximated by a single cavity
resonance, but above threshold and for lower finesse this is not at
all the case \cite{tureci2007,tureci2008}.  The treatment of the
matter equations does not involve the SVEA
and is exactly the same as in Ref. [2] 
, where the key assumption is stationary inversion, which allows the
non-linear  polarization in the Maxwell equation to be replaced by a
non-linear function of the electric field itself.  The new element
is that we keep the second time-derivative of the polarization in
the Maxwell equation and evaluate it by differentiating the
polarization equation. The resulting improved AI/MB equations for
the mode functions $\Psi_\mu$ and the frequencies $k_\mu$ are:
\begin{equation}
\Psi_\mu (\bx) =  \frac{i \g}{\g - i (k_\mu -
k_a)}\frac{k_\mu^2}{k_a^2} \int d\bxp \frac{D_0(\bx)
G(\bx,\bxp;k_\mu)  \Psi_\mu (\bxp)}{\eps(\bxp) (1 + \sum_\nu
\Gamma_\nu |\Psi_\nu (\bxp)|^2)}\, . \label{eqTSG}
\end{equation}
Here $G(\bx,\bxp;k_\mu)$ is the Green function of the open cavity,
$\Gamma_\nu = \Gamma (k_\nu)$ is the gain profile evaluated at
$k_\nu$, $D_0(\bx)= D_0 (1 + d_0(\bx))$ is the spatial pump, and
$\epsilon (\bx) = n^2(\bx)$ is the dielectric function of the cavity
(for the microcavity edge emitting laser $n(\bx) = n_0$ and the pump
is assumed uniform ($d_0(\bx) = 0$)). Electric field and pump
strength are
  measured in units $e_c = \hbar \sqrt{\g
\gpar}/2g, D_{0c}= 4 \pi k_a^2 g^2/\hbar \g$, where $g$ is the
dipole matrix element of the gain medium.  With a slight change in
notation this equation
differs from that derived in Ref. [2] 
only by the additional factor $k_\mu^2/k_a^2$ multiplying the
integral. This is consistent with the expectation that the SVEA is
good when the lasing frequency is very close to the atomic
frequency; incorporating this change into the iterative algorithm
for solving the system is trivial, but crucial for quantitative
agreement with the current MB simulations. However we do not find
qualitative changes from dropping the SVEA, either for this simple
laser or for the more complicated two-dimensional random laser
studied elsewhere \cite{tureci2008}.

In Figs.~1a,~1b we show results for $n_0 = 1.5,3$, finding
remarkable agreement between MB and AI approaches with no fitting
parameters for the case of three mode and four mode lasing
respectively. Both thresholds, modal intensities and frequencies are
found correctly by the AISC approach.  Note that all of these
quantities are direct outputs of the AISC theory, whereas they must
be found by numerical fourier analysis of the MB outputs, which can
introduce some numerical error.  If the earlier AI approach is used
{\it with} the SVEA then significant discrepancies are found, for
example for $n_0=3$ the threshold of the third mode is found to be
higher than from MB while the fourth threshold is too low.  Note
however that the theory with the SVEA does get the right number of
modes and the correct linear behavior for large pumps. We believe
that this original AISC approach {\it does} solve very accurately
the Schr\"odinger-Bloch equations and that the same discrepancies
would arise between MB and SB simulations, although we haven't
confirmed this.

\begin{figure}[htbp]
\centering
\includegraphics[width=7cm]{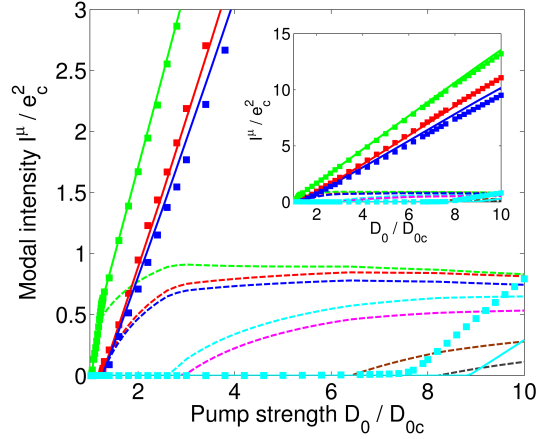}
\caption{Modal intensities as functions of the pump strength $D_0$
in a one-dimensional microcavity edge emitting laser of $n=3$,
$k_aL=20$, $\g=4.0$ and $\gpar=0.001$; the solid lines and data
points are the same as in Fig.~1b.  The dashed lines are the results
of the third order approximation to Eq. (\ref{eqTSG}).  The
frequently used third order approximation is seen to fail badly at a
pump level roughly twice the first threshold value, exhibiting a
spurious saturation not present in the actual MB solutions or the AI
theory. In addition, the third order approximation predicts too many
lasing modes at larger pump strength. For example, it predicts seven
lasing modes at $D_0=10$, while both the MB and AISC show only four.
Right inset just shows the same data on a larger vertical scale.}
\label{fig3rdOrder}
\end{figure}

Almost all studies of the MB equations in the multimode regime have
used the approximation of treating the non-linear modal interactions
to third-order (near threshold approximation) and in fact this
approximation is used quite generally and uncritically throughout
laser theory.  From examination of the form of Eq. (\ref{eqTSG}) it
is clear that it treats the non-linear interactions to all orders,
while the third order treatment would arise from expanding the
denominator to the leading order in $|\Psi_\nu|^2$.   This
third-order version of the AISC theory then becomes similar to
standard treatments of Haken \cite{haken1963,fu1991}, with the
improvement of correctly treating the openness of the cavity and the
self-consistency of the lasing modes in space
\cite{tureci2006,tureci2007}.  An early version of this improved
third order theory was found to have major deficiencies: it
predicted too many lasing  modes and the intensities did not scale
linearly at large pump, but exhibited a spurious saturation
\cite{tureci2006,tureci2005}.  In Fig. \ref{fig3rdOrder} we present
comparisons of the third order approximation to Eq. (\ref{eqTSG}),
which is improved over
Ref. [2] 
because it drops the SVEA and the ``single-pole approximation'' used
there. We find that this improved third order theory still does a
very poor job of reproducing the multimode MB results: it still
predicts too many modes (in this case seven, when there should only
be four at $D_0  =10$ in Fig. \ref{fig3rdOrder}), and shows the same
spurious saturation as found earlier \cite{tureci2006}  because the
third-order approximation cannot give the correct linear behavior
for large pumps. The infinite order treatment of Eq. (\ref{eqTSG})
is both qualitatively and quantitatively essential.

The central approximation required for Eq. (\ref{eqTSG}) is that of
stationary inversion. Previous work by Haken \cite{fu1991} argued
that SIA holds for the MB equations when $\gpar \ll \g, \Delta$ ,
where $\gpar$ is the relaxation rate of the inversion and $\Delta$
is the frequency difference of lasing modes.  For a typical
semiconductor laser $\g \simeq 10^{12} - 10^{13} s^{-1}$ and $\gpar
\simeq 10^8 - 10^9 s^{-1}$, or equivalently $\gpar/\g \simeq 10^{-3}
- 10^{-5}$. For the microcavity edge-emitting laser we are modeling
we took $\g,\Delta \sim 1$ and $\gpar = 10^{-3}$, and we found the
excellent agreement shown in Figs.~1a,~1b. In addition, direct
analysis of the inversion vs. time obtained from the MB simulations
confirm very weak time-dependence in the steady state, justifying
the use of SIA. The previous work did not develop a systematic
theory in which $\gpar/\g, \gpar/\Delta$ appear as small parameters
in the lasing equations, allowing perturbative treatment of
corrections to the SIA; we are now able to do this within the AI
formalism.

Note first that in Eq. (\ref{eqTSG}) the electric field is measured
in units  $e_c \sim \sqrt{\gpar}$ but, unlike $\g$, $\gpar$ does not
appear explicitly.  Hence the solutions of Eq. (\ref{eqTSG}) depend
on
 $\gpar$ only through this scale factor, and Eq. (\ref{eqTSG}) makes the strong
prediction of a universal overall scaling of the field intensities:
$|E(x)|^2 \sim \gpar$ when dimensions are restored.  The
perturbative corrections to Eq. (\ref{eqTSG}) are obtained by
including the leading effects of the beating terms between the
different lasing modes which lead to time-dependence of the
inversion at multiples of the beat frequencies.  These population
oscillations non-linearly mix with the electric field and
polarization to generate all harmonics of the beat frequencies in
principle, but the multimode approximation assumes all the newly
generated fourier components of the fields are negligible. The
leading correction to this approximation is to evaluate the effect
of the lowest sidebands of population oscillation on the
polarization at the lasing frequencies and on the static part of the
inversion, both of which will enter Eq. (\ref{eqTSG}). For
simplicity we present a sketch of the correction to Eq.
(\ref{eqTSG}) in the two-mode regime; details and the
straightforward generalization to more modes will be given
elsewhere.

We start by writing the electric field and the polarization in the
multiperiodic forms $e(\bx,t) = 2 \re{\sum_{\mu=1}^2 \Psi_\mu(\bx)
\exp{(-ik_\mu t)}}$ and $p(\bx,t) = 2 \re{\sum_{\mu=1}^2  p_\mu(\bx)
\exp{(-ik_\mu t)}}$, and allow for the first two side-bands at the
beat frequency $\Delta = k_1 - k_2$ in the inversion, so that the
total inversion is
     $D(\bx,t) = D_s(\bx) + d_{+}(\bx) \exp{(-i\Delta
t)} + d_{-}(\bx) \exp{(+i\Delta t)}$ where the real quantity
$D_s(\bx)$ is the time-independent part of the inversion and
$d_{+}(\bx)  =  d_{-}(\bx)^*$. This ansatz is inserted into the
Bloch equation for the inversion \cite{tureci2006}; solving for the
component of the inversion equation which oscillates at
$\exp{(-i\Delta t)}$ relates $d_+(\bx)$ to the product of the field
and polarization and then substitution of the zeroth order result
for the polarization in terms of the zeroth order static inversion
$D_s^{(0)}$ gives
\begin{equation}
d_+ (\bx) =   \frac{2}{i\hbar} \frac{[\Psi_1 p_2^* - \Psi_2^*
p_1]}{(i \Delta - \gpar)} = \frac{\gpar}{\Delta} f(k_1,k_2)
D_s^{(0)} (\bx)\Psi_1(\bx) \Psi^*_2 (\bx)/e_c^2
\end{equation}
where the dimensionless function $ f(k_1,k_2) = -(i+\Delta/(2\g))/(1
+ \tk_1 \tk_2 - i\Delta/\g), \tk_\nu = (k_\nu-k_a)/\g$ and the
fields are not yet measured in units of $e_c$. The component $d_+$
will mix with the field $\Psi_2$ to yield a contribution to the
polarization $p_1$ at frequency $k_1$, ( and similarly $d_-$ and
$\Psi_1$ mix to contribute to $p_2$),
     \be
p^{(1)}_1(\bx) = \frac{g^2}{i\hbar}\frac{[1 + (\gpar/\Delta)
f(k_1,k_2)| \Psi_2(\bx)|^2/e_c^2]}{\g
-i(k_1-k_a)}\,D_s^{(0)}(\bx)\Psi_1(\bx),\label{eqP1st}
     \ee
where $p^{(1)}_2(\bx)$ is obtained by interchanging subscripts
$1,2$. The correction to the AI formalism is the term in the
numerator explicitly proportional to the small parameter
$\gpar/\Delta$.  However, having found a correction to the
polarizations $p_1,p_2$ we must then evaluate their contribution to
the static inversion. We find
     \be
D_s(\bx) = \frac{D_0}{1 + \sum_\nu \Gamma_\nu |\Psi_\nu (\bx)|^2 +
(\gpar/\g) g(k_1,k_2)|\Psi_1 (\bx)\Psi_2 (\bx)|^2}, \label{eqD1st}
     \ee
where the dimensionless function $g(k_1,k_2) =  (2 + \tk_1^2 +
\tk_2^2) (1 - \tk_1 \tk_2)/[(1 + \tk_1^2)(1 + \tk_2^2)]^2$ and now
and below electric fields have been scaled by $e_c$. Note that the
correction term in Eq. (\ref{eqD1st}) is explicitly proportional to
the second small parameter, $\gpar/\g$. For our simulations $\g
\approx \Delta$ and the functions $f(k_1,k_2),g(k_1,k_2)$ are order
unity. The full correction to the non-linear polarization in Eq.
(\ref{eqP1st}) is obtained by replacing $D_s^{(0)}$ with $D_s$ of
Eq. (\ref{eqD1st}). The corrected polarization leads to corrected
version of Eq. (\ref{eqTSG}) of the AISC theory:
\begin{equation}
\Psi_2 (\bx) =  \frac{i \g D_0}{\g - i (k_2 -
k_a)}\frac{k_2^2}{k_a^2} \int \frac{d\bxp}{\eps(\bxp)}\frac{(1+
\frac{\gpar}{\Delta} f(k_1,k_2)|\Psi_1(\bxp)|^2)\, G(\bx,\bxp;k_2)
\Psi_2 (\bxp)}{ (1 + \sum_\nu \Gamma_\nu |\Psi_\nu (\bxp)|^2 +
\frac{\gpar}{\g} g(k_1,k_2)|\Psi_1 (\bxp)\Psi_2 (\bxp)|^2)}.
\label{eqTSGP}
\end{equation}
$\Psi_1(\bx)$ satisfies the same equation with the subscripts $1$
and $2$ interchanged.

\begin{figure}[htbp]
\centering
\includegraphics[width=7cm]{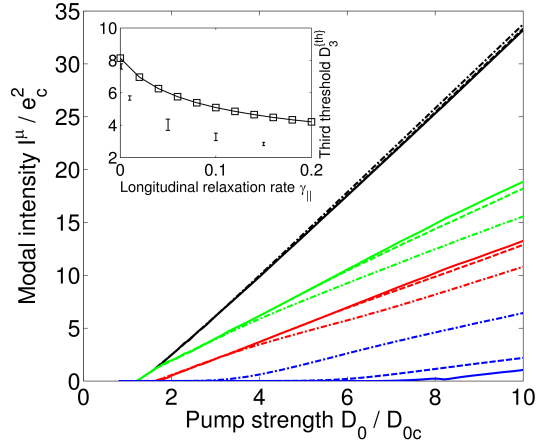}
\caption{Modal intensities for the microcavity edge emitting laser
of Fig.~1a as $\gpar$ is varied ($n=1.5$, $k_aL=20$, $\g=4.0$ and
mode-spacing $\Delta \approx 1.8$. Solid lines are AISC results from
Fig.~1a (with $\gpar=0.001$); dashed lines are $\gpar =0.01$ and
dot-dashed are $\gpar= 0.1$. The color scheme is the same as in
Fig.~1a. The inset shows the shifts of the third threshold as a
function of $\gpar$. The perturbation theory (squares, with the line
to guide the eyes) predicts semiquantitatively the decrease of the
threshold as $\gpar$ increases found in the MB simulations. The MB
threshold is not sharp and we add an error bar to denote the size of
the transition region.} \label{figGammaPar}
\end{figure}

Eq. (\ref{eqTSGP}) predicts corrections to the universal behavior,
$|E(\bx)|^2 \sim \gpar$, found in Fig. \ref{figModalIntensity}.
There is no correction to the first mode below the second threshold
as the correction terms all vanish (there needs to be two modes to
have beats).  However, the theory predicts a non-trivial correction
to the threshold of the second mode.  Note that the correction to
the numerator in Eq. (\ref{eqTSGP}) does not vanish below the second
threshold but contributes self-consistently to that threshold.  This
correction can be regarded as modifying the dielectric function of
the microcavity to take the form $\epsilon'(\bx) = \epsilon(\bx)/[1+
\frac{\gpar}{\Delta}f(k_2,k_1)|\Psi_1(\bx)|^2 ]$; the effective
dielectric function then becomes complex and varying in space
according to the intensity of the first mode.  This in turn changes
the threshold for the second mode.  If the modes are on opposite
sides of the atomic frequency and $k_2 < k_1$, the imaginary part of
the effective index is always amplifying and tends to decrease the
second threshold; we find this effect dominates over the change in
the real part and increasing $\gpar$ uniformly decreases the
thresholds. The opposite effect is possible and observed in other
cases we have studied (not shown). In Fig. \ref{figGammaPar} we show
the results of MB simulations as $\gpar$ is varied from $0.001$ to
$0.1$ (with $\g = 4$).  Note that the universal behavior (in units
scaled by $\gpar$) is well obeyed until $\gpar = 0.1$, encompassing
most lasers of interest.  The qualitative effect predicted by the
perturbation theory is clearly seen, the higher thresholds are
reduced as $\gpar$ is increased. The effect is small for the 2nd
threshold but large for the third as we expect as the corrections
scale with the product of the intensities of lower modes.  The inset
to Fig. \ref{figGammaPar} shows that the perturbation theory for the
third threshold (a suitable generalization of the two-mode Eq.
(\ref{eqTSGP})) yields semiquantitative agreement with the threshold
shifts found from the simulations.  Detailed comparisons between
multimode perturbation theory and simulations above threshold will
be given elsewhere. Note that the simulations also find additional
modes turning on for $\gpar = 0.1$ but their intensities are very
small and they are not shown in Fig. \ref{figGammaPar}.

In conclusion, for the case studied, the recently developed ab
initio self-consistent laser theory without the slowly varying
envelope approximation presents a very accurate solution of the
steady-state Maxwell-Bloch semiclassical lasing equations without
solving the time-dependent problem.  Third order treatments fail
badly and our infinite order treatment is essential. The theory is a
well-controlled expansion in the small parameters
$\gpar/\Delta,\gpar/\g$ and leading corrections in these small
parameters can be evaluated and understood qualitatively.  The AI
method is flexible and more convenient than MB simulations for
complex two-dimensional and three-dimensional microcavities and
provides more physical insight into the lasing properties.  The
accuracy of the method suggests it can be useful in the analysis and
design of novel laser systems.

\section*{Acknowledgments}
We would like to thank Stefan Rotter, Hui Cao and Jonathan Andreasen
for helpful discussions. This work was supported in part by NSF
grant no. DMR-0408636.

\end{document}